\documentclass[aps,twocolumn,prd,showpacs,preprintnumbers,nofootinbib,amsmath,amssymb]{revtex4}

\usepackage{graphicx}
\usepackage{dcolumn}
\usepackage{bm}
\usepackage{amsmath}

\newcommand{\beq}{\begin{eqnarray}}
\newcommand{\eeq}{\end{eqnarray}}

\newcommand{\eg}{{\it e.g.\ }}
\newcommand{\real}{{\sf I}\kern-.12em{\sf R}}
\newcommand{\comp}{{\sf I}\kern-.50em{\sf C}}
\newcommand{\unity}{{\sf I}\kern-.54em{\sf 1}}

\def\spose#1{\hbox to 0pt{#1\hss}}
\def\ltapprox{\mathrel{\spose{\lower 3pt\hbox{$\mathchar"218$}}
 \raise 2.0pt\hbox{$\mathchar"13C$}}}

\begin{document}

\title{The Roberge-Weiss endpoint in $N_f = 2$ QCD.}
\author{Claudio Bonati$^{1}$, Guido Cossu$^2$, Massimo D'Elia$^{3}$ and Francesco Sanfilippo$^{4}$}
\affiliation{$^1$Dipartimento di Fisica, Universit\`a
di Pisa and INFN, Sezione di Pisa, Largo Pontecorvo 3, 56127 Pisa, Italy\\
$^2$ 
Theory Center, IPNS, High Energy Accelerator Research Organization (KEK), Tsukuba
305-0801, Japan\\
$^3$Dipartimento di Fisica, Universit\`a
di Genova and INFN, Sezione di Genova, Via Dodecaneso 33, 16146 Genova, Italy\\
$^4$Dipartimento di Fisica, 
Universit\`a di Roma ``La Sapienza'' and INFN, Sezione di Roma, Piazzale A. Moro 5, 00185 Roma, Italy}

\date{\today}

\begin{abstract}
We present the results of extensive simulations regarding the critical
behavior at the endpoint of the Roberge-Weiss transition for 
$N_f = 2$ QCD. We confirm
early evidence, presented in Ref.~\cite{rwe}, according to which 
the Roberge-Weiss endpoint is first order in the limit of large
or small quark masses, and second order for intermediate masses.
A systematic study of the transition strength as a function of 
the quark mass in the first order
regions, permits us to estimate the tricritical values of the 
quark mass separating
the second order region from the first order ones.
\end{abstract}

\maketitle

\section{Introduction}

A full understanding of the QCD phase diagram 
at finite temperature $T$ and baryon
chemical potential $\mu_B$
is one of the main unreached goals within
the Standard Model of Particle Physics. Various questions
remain open, which are of fundamental importance both theoretically
and phenomenologically, for astrophysics 
and heavy ion collisions, 
like the existence and location of a
possible critical endpoint in the $T-\mu_B$ plane, accessible 
to experiments.

Lattice QCD simulations, which are in principle the ideal tool
for a full non-perturbative investigation of the phase diagram,
are unfortunately hindered at $\mu_B \neq 0$ by the complex nature
of the path integral measure (sign problem). Among other approximate 
methods, a way to partially overcome the sign problem is to consider a
purely imaginary quark chemical potential, $\mu_q \equiv \mu_B/3 = i \mu_{I}$:
numerical simulations are feasible 
and information about real $\mu_B$ can be recovered by 
analytic continuation techniques~\cite{alford,lombardo,muim,muim2,immu_dl,azcoiti,chen,giudice,cea,sqgp,conradi,cea2,sanfo1,cea3,cea4}. 

Recent literature has pointed out that the phase structure at finite
$T$ and imaginary chemical potential may be important by its own,
and teach us something about the non-perturbative properties of QCD
also at zero or small real 
$\mu_B$~\cite{sqgp,Kouno:2009bm,rwe,sakai,rwe2,aarts}. 
Such phase structure is characterized 
by a periodicity of the partition function
\beq
Z(T,\mu_I) = {\rm Tr} \left( e^{-\frac{1}{T}
\left(\hat{\cal H}_{QCD} - i \mu_I \hat N_q \right)}
\right) 
\eeq
in the angular variable $\theta = \mu_I/T$, which can be viewed, in the 
path integral representation of the partition function, as a 
phase rotation of fermion boundary conditions in the Euclidean 
temporal direction. It can be shown~\cite{rw} that 
the period in $\theta$ is $2 \pi /N_c$,
where $N_c$ is the number of colors. Such periodicity is smoothly realized
in the low temperature, confined phase, as expected from the fact that
only uncolored states, with $N_q$ multiple of $N_c$, contribute to the 
system dynamics. 

The situation is different 
in the high temperature phase, as expected from the fact that
also colored states appear. Indeed,
as can be explicitly verified by perturbative 
computations~\cite{rw}, the periodicity is realized in a non-analytic way:
the system goes through first order lines, 
known as Roberge-Weiss (RW) transitions, when $\theta$ crosses 
some fixed values,  $\theta_k = (2 k + 1)\pi/N_c$, where $k$ is an integer.
For such values of $\theta$ the system possesses an exact $Z_2$ symmetry,
which is spontaneously broken for $T > T_{\rm RW}$
and unbroken for $T < T_{\rm RW}$: therefore at $T = T_{\rm RW}$,
which is in fact the endpoint of the RW lines, a genuine finite $T$
phase transition takes
place for all values of the quark masses.
Such transition coincides with the phase
transition at which charge symmetry is spontaneously broken 
when a spatial dimension is compactified below a given
critical size (see e.g. Refs.~\cite{degrand0,degrand,lucini,lucini2,myers} 
for early lattice studies of such transition, which has been investigated
in the context of orientifold planar equivalence 
\cite{asv, uy2006}).

The endpoint of the RW lines has been considered
by recent literature~\cite{sqgp,Kouno:2009bm,rwe,rwe2,aarts,sakai2}, 
for its possible influence on the critical properties
and on the phase diagram of QCD. 
The endpoint can be second order in
the 3D Ising universality class, or first order; in the latter case 
it is actually a triple point, from which two further first order lines
depart.

In Ref.~\cite{rwe} first evidence has been presented showing that, for 
QCD with two degenerate flavors ($N_f = 2$), 
the endpoint is first order in the limit
of small quark masses and second order for intermediate masses;
first order comes back in the high quark mass regime,
where the system reaches its quenched limit.
In the same paper it has been pointed out that, when the 
endpoint is first order (triple point), one of the further first
order lines departing from it can be identified with (part of) the 
continuation of the critical line to imaginary chemical potential,
thus explaining early evidence~\cite{muim,muim2} that the latter 
meets the RW line right on its endpoint. A further conjecture, put
forward in Ref.~\cite{rwe}, has been that the nature of the
transition at $\mu = 0$ as a function of the quark mass spectrum (which
is summarized in the so-called Columbia plot) is regulated by the 
physics of the RW endpoint itself, i.e. that the $\mu = 0$ transition is first
order only when the first order line departing from the RW triple 
point reaches the $\mu = 0$ axis.

Recently the numerical study of the RW endpoint has been extended to $N_f = 3$ 
QCD~\cite{rwe2}, confirming also for this case the presence of a first
order transition for small and high quark masses, with a second
order region for intermediate masses. Moreover, the authors of Ref.~\cite{rwe2}
have suggested that the tricritical behaviour which is present
at the two tricritical masses, separating the second order  
from the first order regions, may shape the critical line also for real 
values of the chemical potential, implying a weakening of the 
transition with real chemical potentials which was suggested also 
by earlier works~\cite{deph1}.

All the results and conjectures above 
claim for a more systematic study of the phase
diagram in the $T - \mu_I$ plane, which is 
perfectly feasible with present simulation algorithms.
The aim of the present work is to move a step in this 
direction, by extending in a substantial way the original results
presented in Ref.~\cite{rwe} for $N_f = 2$ QCD. In particular 
we will present results about the critical behavior at the RW
endpoint for a large set of quark masses, confirming the results
of Ref.~\cite{rwe} and giving an estimate
for the two tricritical masses, $m_{t1}$ and 
$m_{t2} > m_{t1}$, separating the first order regions
from the second order one.

Our first instrument to discern the critical behavior around
the RW endpoint is the finite size scaling of various susceptibilities.
However, an accurate determination of the critical properties around 
the tricritical point may be a non-trivial task. Much can be learned
in this direction by the study of simpler statistical systems, like
the 3D 3-state Potts model
in presence of a negative magnetic field $h$~\cite{negpotts,pottsim}, which shares
some of the properties of QCD along the RW lines, 
i.e. the presence of a residual 
$Z_2$ symmetry which gets spontaneously broken at a critical temperature.
In that model the transition is first order for small values of $|h|$
and second order for large values of $|h|$, with a tricritical value
of the field, $h_{\rm tric}$, separating the two regimes\footnote{
In the Potts model, of course, one does not observe the re-strengthening 
of the transition (hence a second tricritical point), 
which is present for QCD at low masses and which 
is likely caused by the interplay with chiral degrees of freedom.}.
As shown in Ref.~\cite{negpotts},
discerning the correct universality
class close to $h_{\rm tric}$ is difficult since,
at a given distance from $h_{\rm tric}$, tricritical scaling 
will mask the correct critical indexes up to a given lattice size 
$L_{\rm max}$, which is regulated 
by tricritical crossover exponents. A similar phenomenon is expected
around $m_{t1}$ and $m_{t2}$.
Following Ref.~\cite{negpotts}, an alternative strategy will 
be to determine parameters which fix the strength of the first order transition
for $m < m_{t1}$ or $m > m_{t2}$, like the latent heat or the gap
of the order parameter, and extrapolate the values of 
$m$ at which such parameters vanish, i.e. the first order 
transition disappears. 

Our results have been obtained using 
standard rooted staggered fermions on lattices with $N_t = 4$.
The paper is organized as follows: in Sec. II we give more details about
the discretized version of QCD under investigation 
and about the observables and the strategy used
for the study of the critical behaviour; in Sec. III we present our numerical
results and finally, in Sec IV, we discuss our conclusions and perspectives.

\section{Numerical setup}
\label{setup}

We shall consider the partition function of $N_f = 2$ QCD 
in presence of an imaginary chemical potential and
in the standard staggered discretization of dynamical fermions,
\beq
Z(T,\theta) \equiv \int \mathcal{D}U e^{-S_{G}[U]} 
\left( \det M[U,\theta] \right)^{1/2} \, ,
\eeq
where $\theta = \mu_I/T$, $S_G$ is the pure gauge plaquette 
action and $M$ is the fermion matrix
\begin{eqnarray}
M_{i,j} &=& a m
\delta_{i,j} + {1 \over 2} 
\sum_{\nu=1}^{3}\eta_{i,\nu}\left(U_{i,\nu}\delta_{i,j-\hat\nu}-
U^{\dag}_{i-\hat\nu,\nu}\delta_{i,j+\hat\nu}\right) \nonumber \\
&+& \eta_{i,4}
\left(e^{ i a \mu_I}U_{i,4}\delta_{i,j-\hat4}-
e^{- i a \mu_I}U^{\dag}_{i-\hat4,4}\delta_{i,j+\hat4}\right) \, .
\label{fmatrix}
\end{eqnarray}
Here $i$ and $j$ refer to lattice sites, $\hat\nu$ is a unit vector on
the lattice, $\eta_{i,\nu}$ are the staggered phases, 
$a$ is the lattice spacing and
$m$ is the bare quark mass.

RW transitions take place for $\theta = (2 k + 1)\pi/3$.
We shall consider in particular the case $\theta = \pi$: for 
this value the residual $Z_2$ symmetry, which is spontaneously broken 
at $T_{\rm RW}$, corresponds to charge conjugation, hence 
the imaginary part of the Polyakov loop or, alternatively, 
the imaginary part of the baryon number 
can be taken as possible order parameters; as in Ref.~\cite{rwe},
we shall consider the former. In the following $L$ will stand for the
spatially averaged Polyakov loop trace (normalized by $N_c$),
hence ${\rm Im}(L)$ is the order parameter.

The order parameter susceptibility is defined by
\beq
\chi \equiv L_s^3\ (\langle {\rm Im}(L)^2 \rangle - \langle |{\rm Im}(L)| \rangle^2) \, ,
\label{suscdef}
\eeq 
where $L_s$ is the spatial size in lattice units, and
is expected to scale, around the transition, as follows:
\beq
\chi = L_s^{\gamma/\nu}\ \phi (t L_s^{1/\nu}) \, .
\label{fss}
\eeq
where $t = (T - T_{\rm RW})/T_{\rm RW}$ is the reduced temperature.
That means that 
the quantities $\chi/L_s^{\gamma/\nu}$, measured on different lattice sizes, 
should fall on the same curve when plotted against $\tau L_s^{1/\nu}$. 

Another relevant quantity is the specific heat $C$ of the system,
which is instead expected to scale as
\beq
C = C_0 + L_s^{\alpha/\nu}\ \phi_2 (t L_s^{1/\nu}) \, ,  \label{fss2}
\label{Cfss}
\eeq
where $C_0$ is a regular contribution. The values of the critical indexes
$\alpha$, $\gamma$ and $\nu$ which are relevant to our analysis are
listed in Table~\ref{CRITEXP} (see \eg Refs.~\cite{landau, pelissvic}), together with the values they take
for the different critical behaviors which may take place in our system,
i.e. first order, second order in the universality class of the 3D Ising 
model, and tricritical mean field.

A careful verification of Eqs.~(\ref{fss}) and (\ref{fss2}), as well as 
of similar relations giving the finite size scaling behavior of other
relevant quantities, gives information about critical
indexes, hence about the universality class of the transition.
A more direct way, in the
case of a first order transition, is to verify the existence, in 
the thermodynamical limit, of 
finite gaps in the order parameter or in the internal energy (latent heat),
which may be visible by looking at double peak distributions of physical
observables around the transition, or by 
studying the large volume limit of some cumulants.

\begin{table}[bt!]
\begin{tabular}{|c|c|c|c|c|c|}
\hline & $\nu$ & $\gamma$ & $\alpha$ & $\gamma/\nu$ & $\alpha/\nu$\\
\hline $3D$ Ising & 0.6301(4) & $1.2372(5)$ & 0.110(1) & $\sim 1.963$ & $\sim 0.175$ \\
\hline Tricritical & 1/2 & 1 & 1/2 & 2 & 1\\
\hline $1^{st}$ Order & 1/3 & 1 & 1 & 3 & 3\\
\hline
\end{tabular}
\caption{Critical exponents relevant to our analysis.}\label{CRITEXP}
\end{table}

\begin{figure}[h]
\vspace{1cm}
\includegraphics[width=0.45\textwidth]{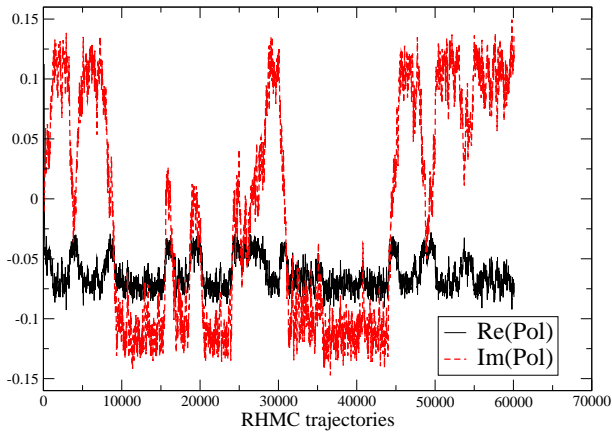}
\caption{Monte-Carlo histories of the real and imaginary part of the Polyakov loop
for a $\beta$ value (5.328) around the critical point and 
$am = 0.0175$ on a $16^3 \times 4$ lattice.
}
\label{fig1}
\end{figure}

\begin{figure}[h]
\vspace{1cm}
\includegraphics[width=0.45\textwidth]{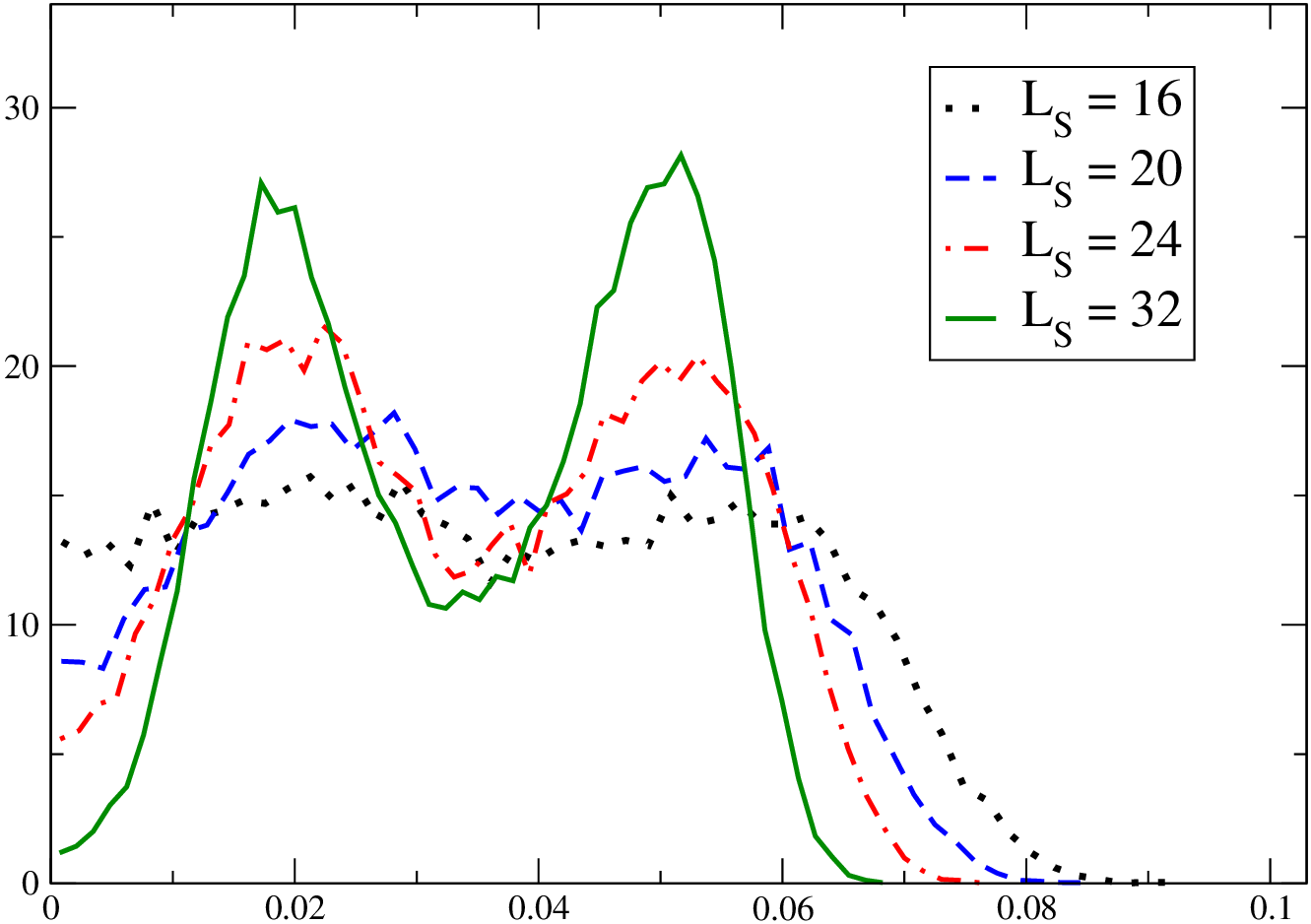}
\caption{Reweighted distribution of the real part of the Polyakov loop at the 
pseudo-critical point for $am = 1.5$ and various lattice sizes.
}
\label{fig3}
\end{figure}

\begin{figure}[h]
\vspace{2cm}
\includegraphics[width=0.45\textwidth]{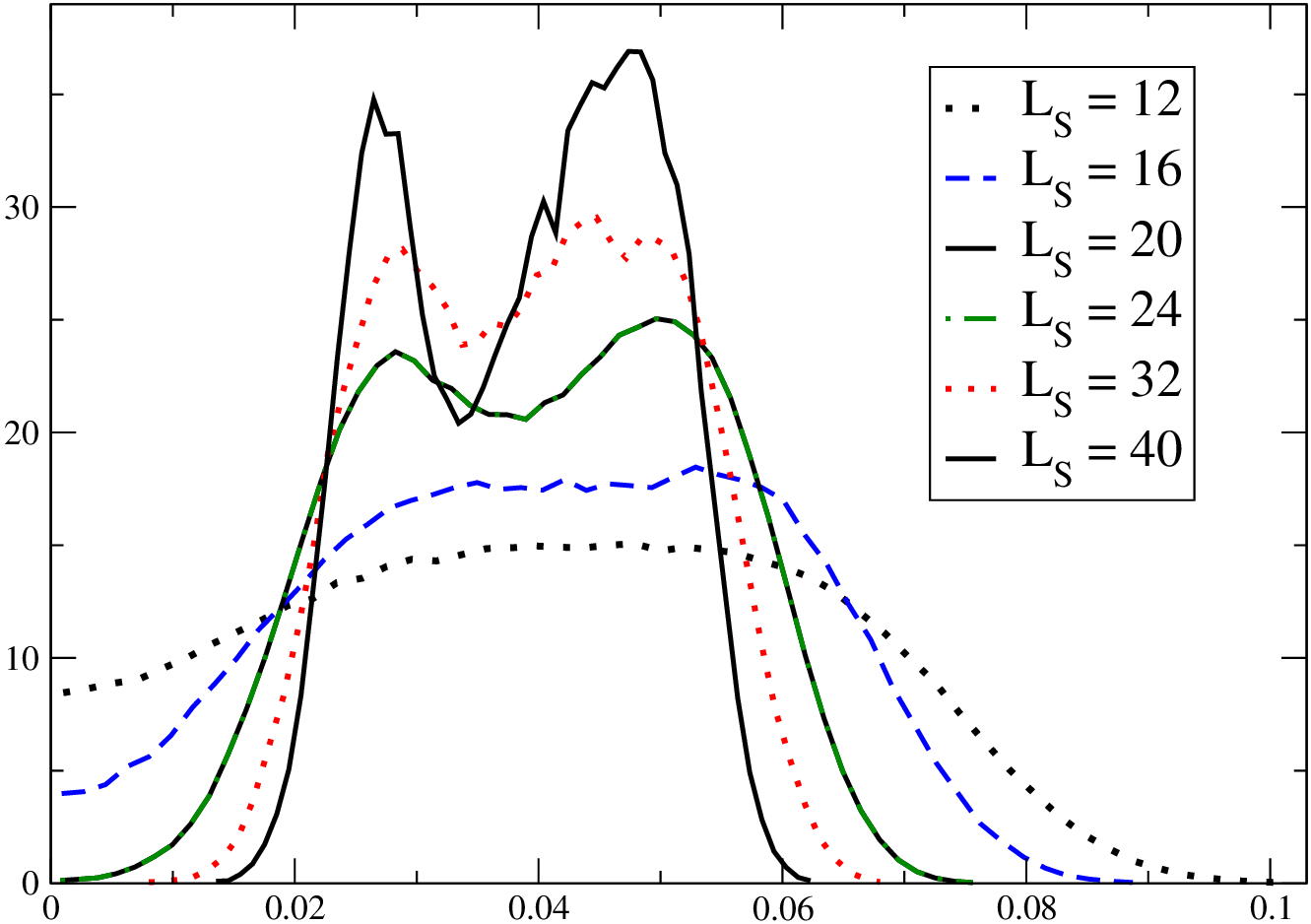}
\caption{Reweighted distribution of the real part of the Polyakov loop at the 
pseudo-critical point for $am = 1.$ and various lattice sizes.
}
\label{fig4}
\end{figure}

\begin{figure}[h]
\vspace{1cm}
\includegraphics[width=0.45\textwidth]{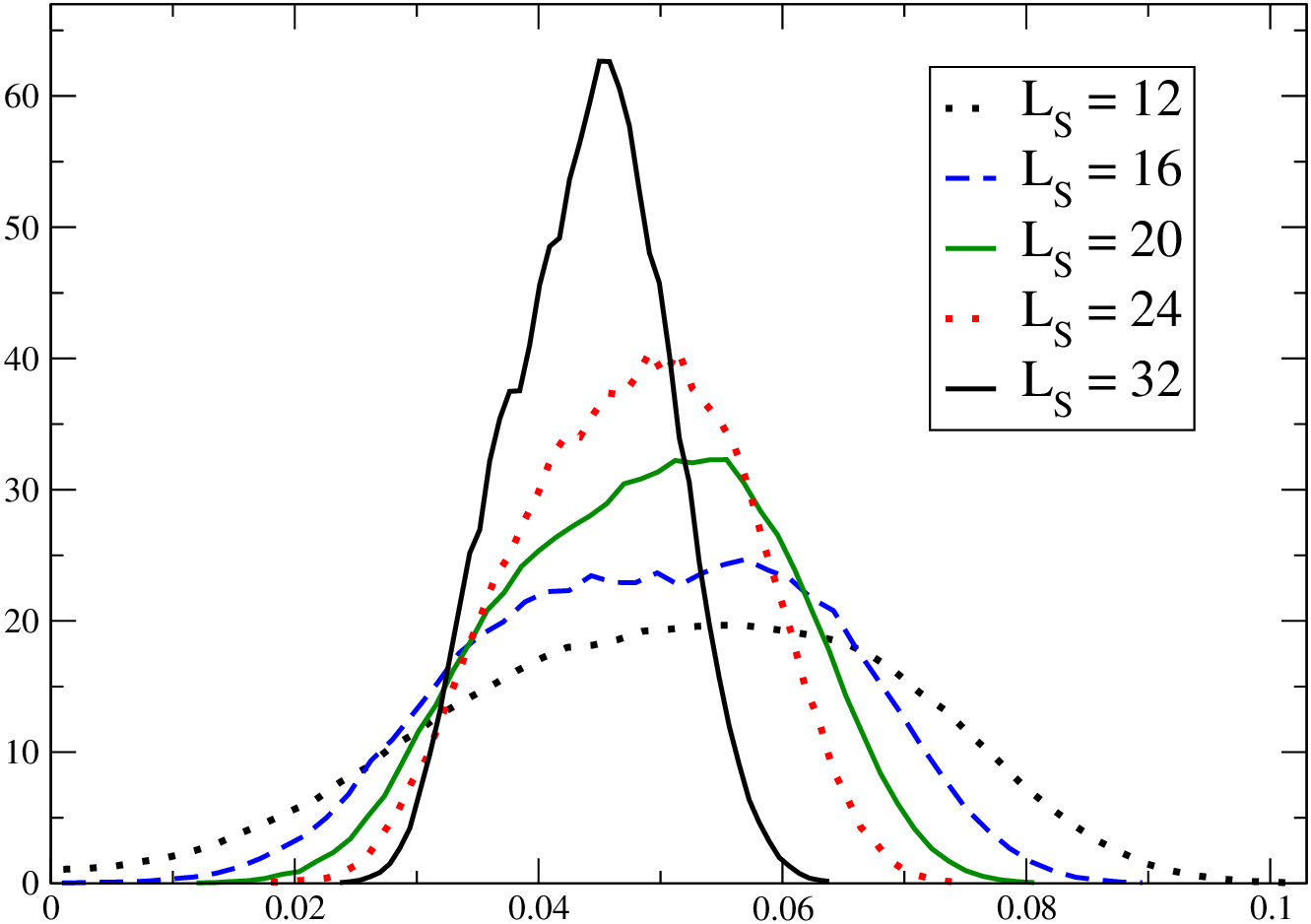}
\caption{Reweighted distribution of the real part of the Polyakov loop at the 
pseudo-critical point for $am = 0.5$ and various lattice sizes.
}
\label{fig5}
\end{figure}

An example is the Binder-Challa-Landau
cumulant~\cite{Challa} of the energy,
which is defined as \(B_4=1-\langle E^4\rangle/(3\langle E^2\rangle ^2)\). It 
can be shown (see \eg~\cite{LeeKosterlitz}) that near a transition \(B_4\) develops minima 
whose depth scales as
\begin{eqnarray}
B_4|_{min} &=&
\frac{2}{3}-\frac{1}{12}\left(\frac{E_+}{E_-}-\frac{E_-}{E_+}\right)^2+O(L_s^{-3}) \nonumber \\
&=&
\frac{2}{3}-\frac{1}{3}\left(\frac{\Delta_E}{\epsilon}\right)^2+O(\Delta_E^3)+O(L_s^{-3})
\label{Bmin}
\end{eqnarray}
where \(E_{\pm}=\lim_{\beta\to\beta_c^{\pm}}\langle E\rangle\), \(\Delta_E=E_+-E_-\) and 
\(\epsilon=\frac{1}{2}(E_++E_-)\). In particular, the thermodynamical limit of  \(B|_{min}\) is less 
than \(2/3\) if and only if a latent heat is present. To simplify our analysis we have
considered the average plaquette (sum of the spatial and temporal plaquettes) in place
of the internal energy, since it is a quantity which can be measured much more easily
and, like the internal energy, is even under the $Z_2$ symmetry which gets broken at 
the RW endpoint. 
To simplify the notation, in the following we will use 
the shorthand 
\begin{equation}
B=\frac{2}{3}-B_4|_{min} \, 
\label{Bdef}
\end{equation}
and from Eq.~(\ref{Bmin}) it follows that $B \propto \Delta_E^2$, where in our case by 
$\Delta_E$ we actually mean the gap at the transition in the average plaquette.

A different, but analogous quantity is the gap of the order parameter, $\Delta$, 
which can be extracted by looking at the scaling of the maximum of its susceptibility, 
$\chi$, and using the relation, valid in the large volume limit for a first order transition,
\beq
\chi_{\rm max} \sim {\rm const.}\ +  \frac{L_s^3}{4} \Delta^2 \, . \label{suscmax}
\eeq

Both $\Delta_E$ and $\Delta$ are expected to vanish as we approach a tricritical mass
$m_{\rm tric}$ from the first order side. In particular, the leading order expected behavior is the following
(see \cite{LawSarb} or \cite{Sheehy} for a brief summary)
\beq
\Delta_E \propto \sqrt{h - h_{\rm tric}} \label{deltae_beh}
\eeq
and
\beq
\Delta \propto \sqrt{|(h - h_{\rm tric})\log(h-h_{\rm tric})|} \label{delta_beh}
\eeq
where we have indicated generically by $h$ the relevant parameter driving
the change from first to second order. It is clear that $h$ is a function
of the quark mass and that close enough to the tricritical point one 
can always set $h - h_{\rm tric} \sim m - m_{\rm tric}$; however, appropriate choices
of $h$ can improve the region around the tricritical mass where 
Eqs.~(\ref{deltae_beh}) and (\ref{delta_beh}) hold. Our choice will be
$h \sim m$ in the low mass region and $h \sim 1/m$ in the high mass region.
It is interesting to notice that Eq.~(\ref{delta_beh}) may seem ambiguous, since a multiplicative redefinition $h \to {\rm const.}\, \times\,  h$
changes the functional dependence; however,
as long as $(h - h_{\rm tric}) \ll 1$, the change is subleading and Eq.~(\ref{delta_beh})
still gives the dominant contribution.

Close to the tricritical points it can be particularly difficult to discern 
the correct critical behavior taking place in the thermodynamical limit. Indeed, while 
first order/3D Ising scaling are expected 
to take place for a continuous range of values of $m$ and exact
tricritical scaling only for specific values $m = m_{\rm tric}$,
what really happens is that tricritical scaling regulates a neighborhood
of $m_{\rm tric}$, whose size goes to zero as $L_s \to \infty$
according to critical indexes known as crossover exponents 
(see \eg \cite{Cardy, BinderDeutsch, pelissvic}). 
Indeed, the true critical behavior of the system can be seen only for $|t|\lesssim |h - h_{\rm tric}|^{1/\phi}$, 
where $t$ is the 
reduced temperature and $\phi$ is the 
crossover exponent, which is by definition $\phi=y_h/y_t$ ($y_t$ and $y_h$ 
are the renormalization group eigenvalues of the relevant variables $t$ and $h - h_{\rm tric}$), in particular 
$\phi=1/2$ in our case~\cite{LawSarb}. Putting the question the other way around, 
on a finite lattice of typical size $L_s$,
$|t|$ can be traded for $L_s^{-1/\nu}$ and the previous condition becomes 
$L_s\gtrsim |h-h_{\rm tric}|^{-\nu/\phi}$; in particular, according to the known tricritical 
indexes in Table~\ref{CRITEXP},
one expects tricritical behavior to dominate and mask the correct thermodynamical limit 
up to a critical size 
\beq
L_c \simeq A\ |h - h_{\rm tric}|^{-1} \, ,
\label{critsize}
\eeq
where $A$ is some unknown 
constant.
Such a behavior has been studied and verified
quantitatively in Ref.~\cite{negpotts} in the case of the 3D 3-state Potts model in a negative
external field, which shares part of the symmetries studied in the present
work.

\begin{figure}[t!]
\vspace{1cm}
\includegraphics[width=0.45\textwidth]{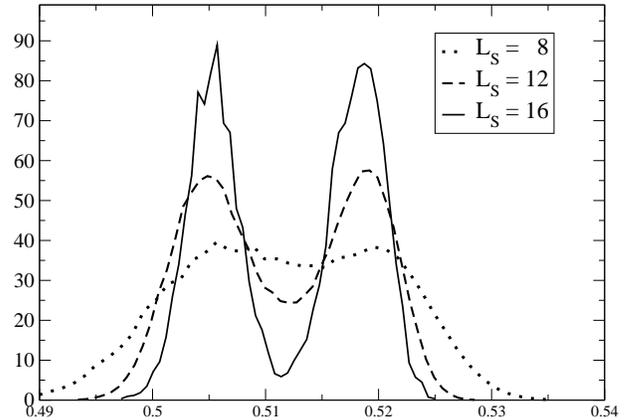}
\caption{Reweighted distribution of the plaquette (average of spatial and temporal) at the 
pseudo-critical point for $am = 0.005$ and various lattice sizes.
}
\label{fig6}
\end{figure}

\begin{figure}[h]
\vspace{1cm}
\includegraphics[width=0.45\textwidth]{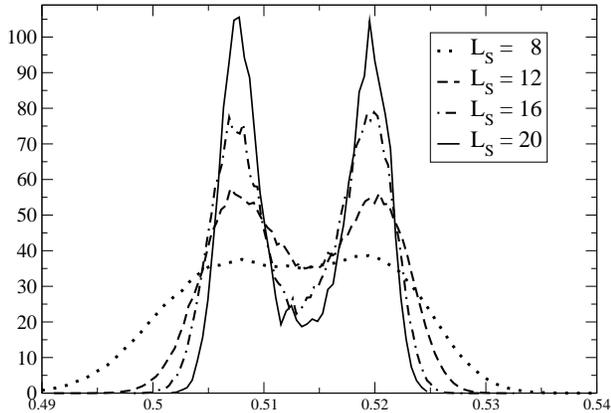}
\caption{As in Fig.~\ref{fig6}, for $am = 0.01$.
}
\label{fig7}
\end{figure}

\begin{figure}[h]
\vspace{1cm}
\includegraphics[width=0.45\textwidth]{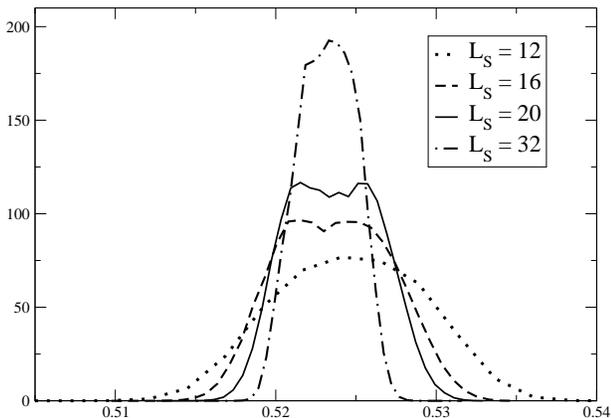}
\caption{As in Fig.~\ref{fig6}, for $am = 0.075$.
}
\label{fig7bis}
\end{figure}

\begin{figure}[t]
\vspace{1cm}
\includegraphics[width=0.45\textwidth]{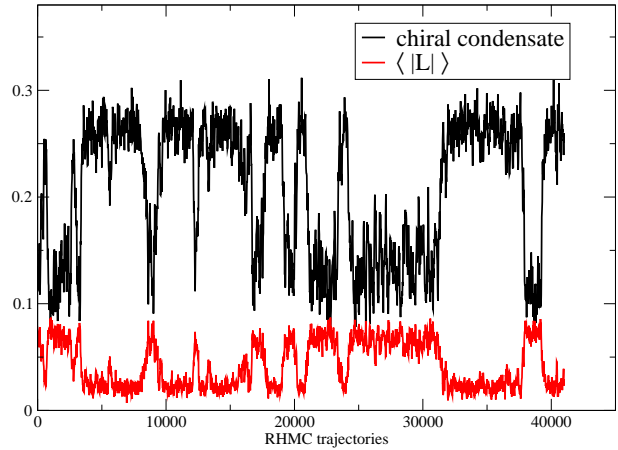}
\caption{Monte-Carlo histories of the Polyakov loop (absolute value)
and of the chiral condensate
for a $\beta$ value (5.314) around the critical point and 
$am = 0.01$, on a $16^3 \times 4$ lattice.
}
\label{fig9}
\end{figure}

\begin{figure*}[t!]
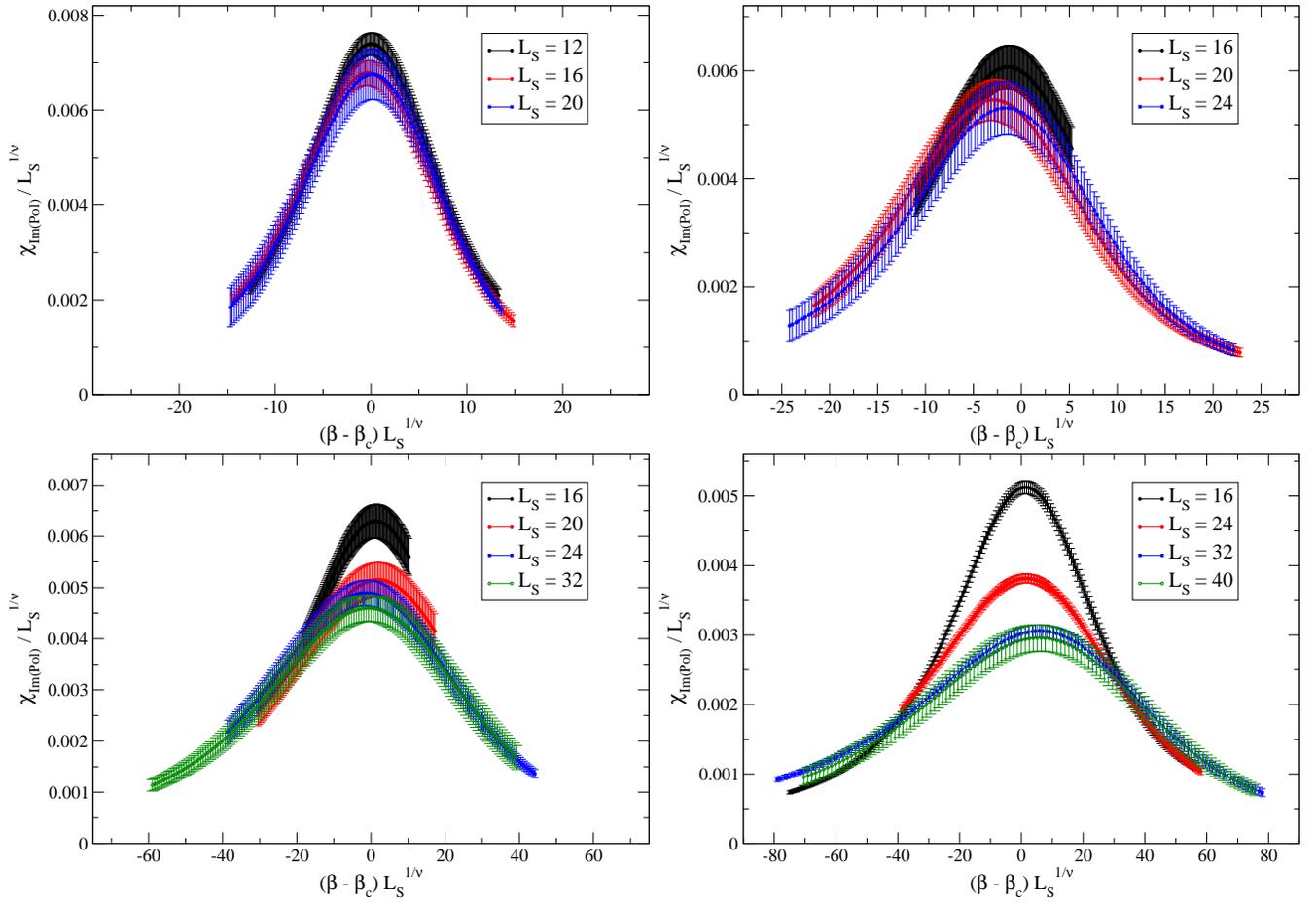

\includegraphics*[width=0.49\textwidth]{scaling_0.0175.eps}
\includegraphics*[width=0.49\textwidth]{scaling_0.030.eps}\\
\includegraphics*[width=0.49\textwidth]{scaling_1.50.eps}
\includegraphics*[width=0.49\textwidth]{scaling_1.00.eps}
\caption{Scaling of the reweighted susceptibility of the imaginary part 
of the Polyakov loop according to first order critical indexes
for $am = 0.0175$ (up-left), $am = 0.03$ (up-right), 
$am = 1.5$ (down-left) and $am = 1.0$ (down-right).} 
\label{fig10a}
\end{figure*}

\begin{figure*}[t!]
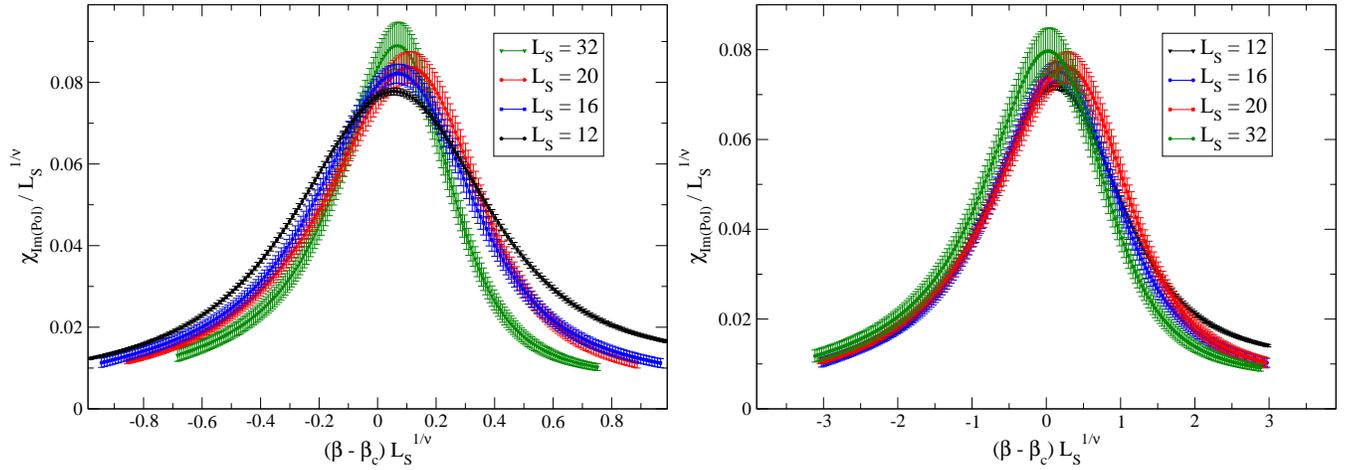

\includegraphics*[width=0.49\textwidth]{scaling_0.075_2nd.eps}
\includegraphics*[width=0.49\textwidth]{scaling_0.075_mf.eps}
\caption{
Scaling of $\chi$ 
for $am = 0.075$ according to 3D Ising critical indexes (left)
and to tricritical mean field indexes (right).}
\label{fig10}
\end{figure*}

The difficulties in discerning the correct critical behavior around 
$m_{\rm tric}$ 
may result in a
difficult determination of 
the tricritical mass itself. For this reason we have followed the 
strategy adopted in Ref.~\cite{negpotts}, i.e. to determine 
the cumulant of the plaquette $B$ and the gap of the order parameter $\Delta^2$
for values of $m$ where a first order transition is present, and then to determine 
$m_{\rm tric}$ by fitting data with the expected behaviors reported in 
Eqs.~(\ref{deltae_beh}) and (\ref{delta_beh}).

With the aim of determining the tricritical masses $m_{t1}$ and $m_{t2}$ present in
the low and high mass regions respectively, we have studied the critical behavior
of the system for various quark masses, 
$am = 0.005, 0.01, 0.0175, 0.025, 0.03, 0.075, 0.2, 0.5, 1., 1.25, 1.5$ and $2.0$.
For each quark mass we have made simulations on lattices with $N_t = 4$ and different
spatial sizes $L_s$, reaching up to $L_s = 40$ when necessary to correctly discriminate
the critical behavior. Numerical simulations have been performed using the standard
Rational Hybrid Monte-Carlo algorithm~\cite{RHMCALG}. Collected statistics have been typically
of the order of $10^5$ trajectories around the critical $\beta$ and for each
value of $L_s$. 

Apart from results obtained for $am = 0.025$ and $am = 0.075$, which 
were already partially reported in Ref.~\cite{rwe}, most numerical simulations have been
performed on two GPU farms located in Pisa and Genoa and provided by INFN, consisting
of a total of 8 S1070 (32 C1060) NVIDIA GPUs. The numerical code, which runs
almost entirely on the GPUs, has been described in detail in Ref.~\cite{gpupaper}.

\section{Numerical results}

\begin{table}
\begin{tabular}{|l|l|l|}
\hline 
\rule{0mm}{3.4mm}\hfill $a m$ \hfill{}     & \hfill $B$ \hfill{} & \hfill $\Delta^2/4$ \hfill{} \\ \hline 
\rule{0mm}{3.4mm}  $0.005$   & $2.15(10)\times 10^{-4}$  &  $9.60(20)\times 10^{-3}$   \\ \hline
\rule{0mm}{3.4mm}  $0.010$  & $1.54(7)\times 10^{-4}$  &  $8.04(26)\times 10^{-3}$   \\ \hline
\rule{0mm}{3.4mm}  $0.0175$   & $1.01(8) \times 10^{-4}$  &  $6.40(40) \times 10^{-3}$   \\ \hline
\rule{0mm}{3.4mm}  $0.025$  & $0.69(4)\times 10^{-4}$  &  $5.54(24) \times 10^{-3}$   \\ \hline
\rule{0mm}{3.4mm}  $0.030$  & $0.48(7) \times 10^{-4}$  &  $4.60(50) \times 10^{-3}$   \\ \hline
\rule{0mm}{3.4mm}  $0.035$  & $0.32(6) \times 10^{-4}$  &  $3.60(40) \times 10^{-3}$   \\ \hline
\rule{0mm}{3.4mm}  $1.00$   & $0.38(4)\times 10^{-5}$  &  $2.59(13)\times 10^{-3}$   \\ \hline
\rule{0mm}{3.4mm}  $1.25$  & $0.58(7)\times 10^{-5}$  &  $4.16(36)\times 10^{-3}$   \\ \hline
\rule{0mm}{3.4mm}  $1.50$   & $0.66(7) \times 10^{-5}$  &  $4.32(24) \times 10^{-3}$   \\ \hline
\rule{0mm}{3.4mm}  $2.00$  & $0.89(7)\times 10^{-5}$  &  $5.20(20) \times 10^{-3}$   \\ \hline
\end{tabular}
\caption{Estimated values for the thermodynamical limit of \(B\) and \(\Delta^2/4\) for values
of the quark mass where a first transition takes place.}\label{table_gap}
\end{table}

\begin{figure}[h!]
\vspace{1cm}
\includegraphics[width=0.45\textwidth]{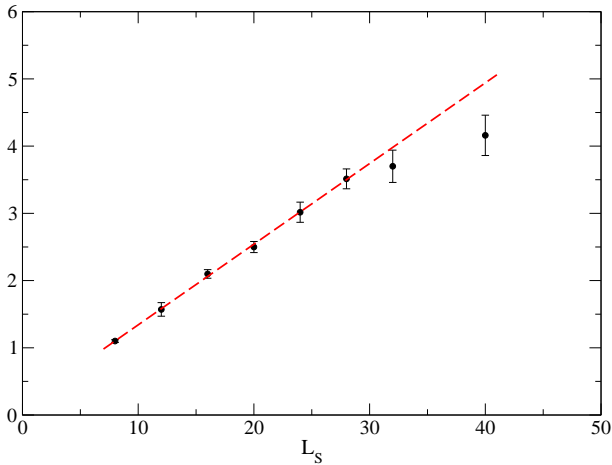}
\caption{Maximum of the susceptibility of the real part of the Polyakov loop as a function 
of the lattice size $L_s$ and for $am = 0.2$, together with a linear fit including sizes $L_s < 32$.
}
\label{m0.2}
\end{figure}

\begin{figure}[t!]
\vspace{1cm}
\includegraphics[width=0.45\textwidth]{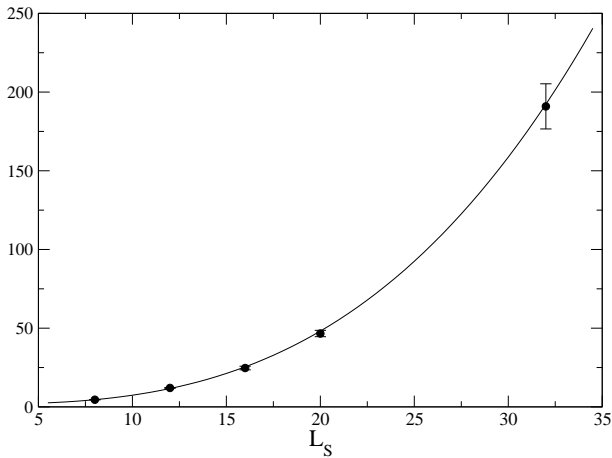}
\caption{Maximum of the susceptibility of the order parameter, $\chi$, as a function
of the lattice size $L_s$ for $am = 0.025$, together with a cubic fit $\chi = {\rm const.}\ +\ \Delta^2 \, L_s^3/4$ including
all sizes ($\chi^2/{\rm d.o.f.} = 0.89$).
}
\label{fig12}
\end{figure}

\begin{figure}[t!]
\vspace{1cm}
\includegraphics[width=0.45\textwidth]{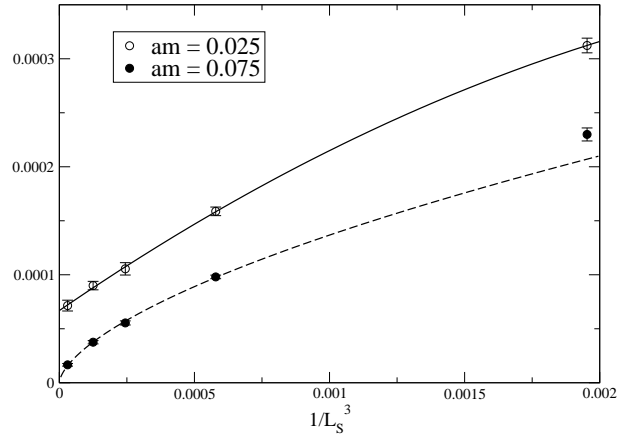}
\caption{Binder-Challa-Landau cumulant of the plaquette 
(see definition in Eq.~(\ref{Bdef})) as a function of the lattice
size for $am = 0.025$ and
$am = 0.075$. In the first case a function $B = a + b/L_s^3 + c/L_s^6$ 
describes well all data with $a =  0.69(4)\times 10^{-4}$ and 
$\chi^2/{\rm d.o.f.} = 0.13$. For $am = 0.075$, instead, 
data with $L_s > 8$ are well described 
($\chi^2/{\rm d.o.f.} = 0.69$)
by a dependence 
$B = a L_s^b$ ($b = 0.62(2)$) which gives $B = 0$ in the thermodynamical limit.
}
\label{fig13}
\end{figure}

\begin{figure}[h]
\vspace{1cm}
\includegraphics[width=0.45\textwidth]{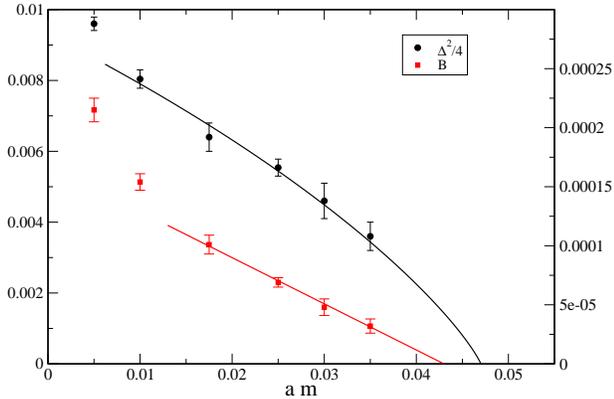}
\caption{Binder-Challa-Landau cumulant of the plaquette, extrapolated
to the thermodynamical limit, and $\Delta^2/4$ for small quark masses where a first order transition
is present. We include the result from a linear fit 
$B_\infty = b\  (a m_{t1} - am)$,
giving the value of the tricritical mass $a m_{t1} = 0.0428(24)$
and $\chi^2/{\rm d.o.f.} = 0.13$ 
(we have included quark masses $am \geq 0.0175$),
and from a fit to Eq.~(\ref{delta_beh}),
$\Delta^2/4 = c\ (a m_{t1} - am) \log (a m_{t1} - am)$,
giving $a m_{t1} = 0.0477(23)$ and $\chi^2/{\rm d.o.f.} = 0.37$
(we have included quark masses $am \geq 0.01$).
}
\label{fig14}
\end{figure}

\begin{figure}[h]
\vspace{1cm}
\includegraphics[width=0.45\textwidth]{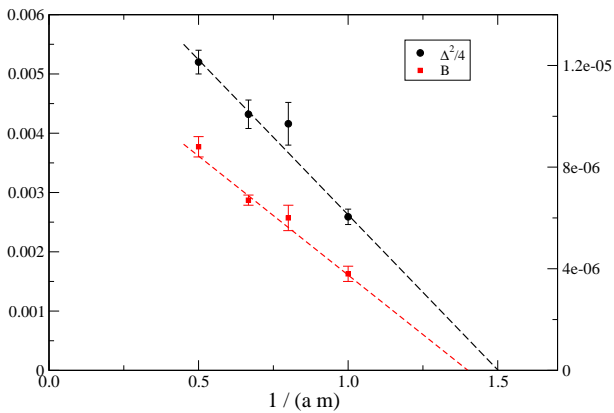}
\caption{Binder-Challa-Landau cumulant of the plaquette, extrapolated
to the thermodynamical limit, and $\Delta^2/4$ for high quark masses 
where a first order transition
is present. We include the result from linear fits 
$B_\infty = b\ (1/(a m_{t2}) - 1/(am))$,
giving $a m_{t2} = 0.71(4)$
($\chi^2/{\rm d.o.f.} = 1.09$), and 
$\Delta^2/4 = c\ (1/(a m_{t2}) - 1/(am))$,
giving $a m_{t2} = 0.67(3)$ ($\chi^2/{\rm d.o.f.} = 1.0$).
All masses have been included in the fit in both cases.
}
\label{fig16}
\end{figure}

The presence of a first order RW endpoint, i.e. of a triple point at the end
of the RW lines, has clear signatures in the Monte-Carlo (MC) histories and in the 
probability distributions of the order parameter and of other quantities. 
In Fig.~\ref{fig1} we show the MC histories of the real and imaginary part
of the Polyakov loop for $am = 0.0175$, 
where the endpoint is first order, and a $\beta$ value around the transition.
Metastabilities are clearly detectable, with ${\rm Im}(L)$, the order parameter,
 taking three distinct possible values, one in the unbroken and two in the broken $Z_2$ phase.
${\rm Re}(L)$, which is $Z_2$ even, takes instead only two distinct values corresponding
to the broken and unbroken phase.

In Figs.~\ref{fig3}, \ref{fig4} and \ref{fig5} we show the reweighted 
distribution of ${\rm Re} (L)$, at the pseudocritical values of 
$\beta$ taking place on the different lattice sizes,
for three values of $am$ in the heavy quark region, $am = 1.5, 1.0$ and $0.5$ respectively.
For $am = 1.5$ and $am = 1.0$ a double peak distribution clearly develops and deepens as 
$L_s \to \infty$, indicating a first order transition, even if in the latter case
one has to reach $L_s = 40$ to clarify the behavior, indicating that in this case
the first order transition is weaker. For $am = 0.5$, instead, the distribution stays
single peaked for all explored volumes, suggesting that the endpoint may be second 
order in this case: this hypothesis is indeed consistent with the determination
of $a m_{t2}$ presented later.

Similar considerations can be made for the light mass region. 
In Figs.~\ref{fig6}, \ref{fig7} and \ref{fig7bis} we show the reweighted 
plaquette distributions at the pseudocritical couplings for 
$am = 0.005, 0.01$ and $0.075$ respectively. Double peak distributions
are present for the two lower masses, with the first order being clearly
stronger for $am = 0.005$. For $am = 0.075$ instead, as already shown in 
Ref.~\cite{rwe}, the distribution stays single peaked, suggesting that
the endpoint is second order in this case: this is consistent with our 
determination of $a m_{t1}$ (see later).

It is interesting to notice that, when the transition is first order, a 
gap develops also in other quantities, including the chiral condensate,
as visible from Fig.~\ref{fig9}, where we show the MC histories 
of the chiral condensate and of the Polyakov loop around the RW endpoint.
That suggests that, as for the usual thermal transition at $\mu = 0$,
a strict correlation between deconfinement and chiral symmetry restoration
may be present also at the RW endpoint.

These results already fully confirm the outcome of Ref.~\cite{rwe}: the RW endpoint is first 
order in the chiral limit and weakens as the quark mass is increased, till an intermediate
mass region is reached where the transition is second order; it is first order
again in the high quark mass limit, where it weakens as the quark mass is decreased. Last result
is in some sense trivial since, as already discussed in Ref.~\cite{rwe},
it is expected from the fact that the $SU(3)$ pure gauge
transition is first order.

Further confirmations come from looking at the finite size scaling of the susceptibility of 
the order parameter, $\chi$, which is shown in Fig.~\ref{fig10a} 
for $am = 0.0175, 0.03, 1.5$ and $1$. 
The first order scaling ansatz, Eq.~(\ref{fss}), is always verified for the largest
volumes available. However, typically one has to go beyond some critical
size before seeing the correct asymptotic critical behavior, and this critical size 
increases as the transition weakens, i.e. as we approach the tricritical points.
For instance, at $am = 1$ first order scaling sets in only for $L_s \geq 32$.

Similar considerations apply to the second order region. On the 
left-hand side of Fig.~\ref{fig10},
which is taken from Ref.~\cite{rwe}, we show the finite size scaling 
of $\chi$ for $am = 0.075$ according to 3D Ising critical indexes: scaling
is fair for the heights of the peaks and less fair for the widths. On the contrary,
we realize that tricritical mean field indexes perform much better, as apparent
from the right-hand side of Fig.~\ref{fig10} (notice from Table~\ref{CRITEXP} 
that $\gamma/\nu$, regulating the height of the peaks, is practically the same
for 3D Ising and tricritical mean field, while $1/\nu$, which regulates the
widths of the peaks, is different). That does not mean, of course, that 
$am = 0.075$ is exactly equal to one of the two tricritical masses, but rather
that it is close enough to one of them so that a fake tricritical scaling masks
the correct asymptotic scaling at least for sizes up to $L_s = 32$. However, we do 
not know neither how close we are to the tricritical mass, nor how large we have to 
go with $L_s$ to reach the thermodynamical limit, since we have no apriori knowledge of the 
prefactor appearing in Eq.~(\ref{critsize}).

A quantity which is well suited for discerning 3D Ising from tricritical behavior is the specific
heat $C$. Indeed the coefficient $\alpha/\nu$, which regulates the scaling of the height of the singular
part of $C$ (see Eq.~(\ref{Cfss})), changes appreciably when going from tricritical to 3D Ising critical behavior
(see Table~\ref{CRITEXP}), hence deviations from tricritical scaling are expected to appear first in
such quantity. A direct measure of the specific heat of the system is not an easy task, however the susceptibility
of any quantity, 
sharing the same transformation properties of the energy under the relevant $Z_2$ symmetry,
is expected
to scale in the same way: examples are given by the plaquette or by the real part of 
the Polyakov loop, which are both $Z_2$ even.
In Fig.~\ref{m0.2} we show the susceptibility of the real part of the Polyakov loop 
as a function of $L_s$ for $am = 0.2$, which we expect to be in the 3D Ising region. 
It is apparent that data follow a linear behavior (i.e. $\alpha/\nu = 1$), with deviations visible only for
$L_s \geq 32$ and going in the direction of a smaller value of $\alpha/\nu$ (as expected for 3D Ising); 
in particular
in the figure we have plotted the result from a linear fit 
to data up to $L_s = 28$.

Therefore, in order to get a more reliable determination of the tricritical masses, we follow
the strategy described in Sec.~II and proceed
to a determination of the gap of the order parameter and of the plaquette as a function of the 
quark mass in the first order regions.
In Fig.~\ref{fig12} we plot the maxima of the order parameter susceptibility, $\chi$, as a function
of $L_s$, for $am = 0.025$, together with a fit to the asymptotic expected behavior,
Eq.~(\ref{suscmax}), from which we extract $\Delta^2/4$, The same procedure has been repeated 
for all quark masses where a first order transition is present. In Fig.~\ref{fig13}, instead,
we plot the Binder-Challa-Landau cumulant of the plaquette, $B$ (see Eq.~(\ref{Bdef})), as a function of $1/V$
for $am = 0.025$ and $am = 0.075$: in the first case the cumulant extrapolates to a non-zero value
as $V \to \infty$, with both linear and quadratic corrections in $1/V$ clearly visible, while 
in the second case data are well described by a power law and $B = 0$ as $V \to \infty$, indicating
the absence of a gap in the plaquette.

In Table~\ref{table_gap} we summarize all determinations
obtained for $B$ and $\Delta^2/4$. From such values we can try 
to determine the tricritical masses as the points where $B$
and $\Delta$ vanish, fitting data to the expected behaviors 
shown in Eqs.~(\ref{deltae_beh}) and (\ref{delta_beh}). In Fig.~\ref{fig14}
we show the results of such fits in the low mass region
for $B$ and $\Delta^2/4$, respectively. We obtain
$a m_{t1} = 0.0428(24)$ from $B$. Instead, from 
$\Delta^2/4$, we get  $a m_{t1} = 0.0477(23)$ if
we fix $h = m$ in Eq.~(\ref{delta_beh}), however
in this case one should take into account also the systematic 
uncertainty related to a possible multiplicative redefinition, $h =  A_h m$.
In order to further check that our results for $B$ and $\Delta^2/4$ can indeed be described
in terms of a common tricritical mass, we have also performed a combined fit to all data obtained
in the low mass region according to
\beq
B &=& b\ (am - am_{t1}) \nonumber \\
\Delta^2/4 &=& c\ (am - am_{t1})\log(A_h (am-am_{t1})) \, ;\label{combined}
\eeq
including directly, in this case, the possible multiplicative redefinition $h =  A_h m$
among the fit parameters. The best fit gives $b = -387(46)$,
$c = 0.17(6)$, $A_h = -9(5)$ and $m_{t1} = 0.043(2)$, with a $\chi^2/{\rm d.o.f.} = 0.3/4$:
the hypothesis is therefore well verified, but
we cannot trust the uncertainties on the parameters deriving by this best fit, since 
data for $B$ and $\Delta^2/4$ are correlated; notice also that
the multiplicative constant $A_h$ is poorly determined. Staying conservative with
the error estimate, we take as our final determination $m_{t1} = 0.043(5)$.

In Fig.~\ref{fig16} we show instead the same kind of fits
for the high mass region: in this case we have used $1/ (a m)$ as the relevant
variable $h$, as explained in Sec.~II. We obtain 
$a m_{t2} = 0.71(4)$ from $B$.
Instead, regarding $\Delta^2$, we notice that
$(h - h_{\rm tric})$ is $O(1)$ and it makes no sense to 
look for logarithmic corrections (see Eq.~(\ref{delta_beh})): 
a simple linear fit for $\Delta^2$ (see Fig.~\ref{fig16}) gives
$a m_{t2} = 0.67(3)$. However, also in this case
we can redefine $h = A_h/m$ and try again a combined fit according to
\beq
B &=& b\ \left(\frac{1}{am} - \frac{1}{am_{t2}} \right) \\
\Delta^2/4 &=& c\ \left(\frac{1}{am} - \frac{1}{am_{t2}}\right)
\log\left(A_h \left(\frac{1}{am}-\frac{1}{am_{t2}}\right)\right) \, ; \nonumber \label{combined2}
\eeq
leading to $a m_{t2} = 0.72(5)$ and $A_h \sim 10^{-2}$, with  $\chi^2/{\rm d.o.f.} = 2.2/4$.
Also in this case one should take into account correlations among data for  
$B$ and $\Delta^2/4$, hence we prefer to stay conservative in our error estimate and 
state $a m_{t2} = 0.72(8)$.

We notice that both determinations, $a m_{t1} = 0.043(5)$ and 
$a m_{t2} = 0.72(8)$, 
are consistent with the fact that the quark masses for which
no metastabilities and double peak distributions are observed ($am = $ 0.075, 0.2, 0.5) 
are within the second order region.

\section{Conclusions and Perspectives}

\begin{figure}[]
\vspace{1cm}
\includegraphics[width=0.45\textwidth]{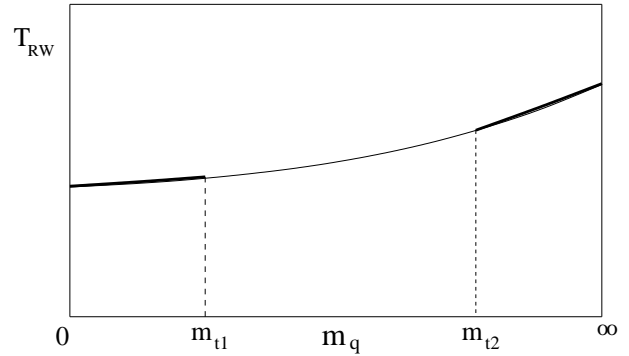}
\caption{Sketch of the phase diagram in the $T$-$m_q$ plane which summarizes 
our results: in $N_f = 2$ QCD the endpoint of the Roberge-Weiss transition is first order
close to the chiral and to the quenched limit and second order for intermediate
masses. A conservative estimate for the two tricritical masses separating the second order region from
the first order ones, for the lattice discretization adopted
in the present work, is $a m_{t1} = 0.043(5)$ and $a m_{t2} = 0.72(8)$. 
}
\label{fig18}
\end{figure}

We have confirmed the outcome of Ref.~\cite{rwe} regarding
the order of the endpoint of the RW transition for $N_f = 2$ QCD: a first order endpoint (triple point)
is present both in the low mass and in the high mass limit; the endpoint is second 
order for intermediate quark masses, which are separated from the first order regions by 
two distinct tricritical masses. Following an investigation performed
in Ref.~\cite{negpotts} for the 3D 3-state Potts model in a negative
external field, which shares part of the same symmetries studied in the present
work, we have performed
a careful study of some parameters directly linked to
the strength of the first order transition, in particular the
Binder-Challa-Landau cumulant of the plaquette and the gap of the order parameter; that
has permitted to obtain independent and consistent determinations of the two
tricritical masses. Staying conservative with error estimates, we state
as our final result 
$a m_{t1} = 0.043(5)$ and $a m_{t2} = 0.72(8)$.
Such results are summarized in Fig.~\ref{fig18}, where we sketch 
a phase diagram in the $T$-$m_q$ plane.

The value of $a m_{t1}$ corresponds to a pion mass of the order of 400 MeV, hence we conclude
that for physical quark masses the RW endpoint should be well inside the first order region.
It is therefore of primary importance to explore what is the fate of the further first order 
lines departing from the triple point. One of them, in particular,  
may reach the zero density axis or have a critical endpoint arbitrarily close to it,
which could have great influence on the physics of strongly interacting matter right above the 
deconfinement transition. The question is also strictly connected to the problem of the 
order of the chiral transition for $N_f = 2$~\cite{2flv1,2flv2}.

 Another important issue is of course to extend our investigation 
to $N_f \neq 2$ and confirm the conjecture that the nature of the
transition at $\mu = 0$ may be regulated by the physics of the RW endpoint~\cite{rwe},
i.e. that the $\mu = 0$ transition is first
order only when the first order line departing from the RW triple 
point reaches the $\mu = 0$ axis, and that tricritical scaling 
indeed shapes the chiral critical surface~\cite{rwe2}.

All these investigations will require extensive numerical simulations,
which are however perfectly feasible since they involve an imaginary chemical
potential. Part of this program is progress.

We stress that our present results are valid for the standard rooted staggered discretization of the theory and for 
lattices with $N_t = 4$, corresponding to a lattice spacing of about $0.3$ fm. A key issue is then 
also to verify that the main features of the phase diagram remain unchanged when 
changing discretization and/or approaching the continuum limit. The two tricritical masses could 
still be present, but the first order regions could in principle extend or shrink in a significant
way.

\section*{Acknowledgments}

We thank Ph.~de Forcrand, A.~Di Giacomo, O.~Philipsen and E.~Vicari for useful discussions.

\end{document}